# A new tool in nuclear physics: Nuclear lattice simulations

Ulf-G. Meißner [1,2,3,†]

[1]*Helmholtz-Institut für Strahlen- und Kernphysik and BCTP, Universität Bonn, Nußallee 14-16, D-53115 Bonn, Germany*
[2]*Forschungszentrum Jülich, IAS-4, IKP-3, JCHP and JARA HPC, D-52425 Jülich, Germany*
[3]*Kavli Institute for Theoretical Physics China, CAS, Beijing 100190, China*

[†]*Electronic address:* meissner@hiskp.uni-bonn.de

In the last years, chiral effective field theory has been successfully developed for and applied to systems with few nucleons. Here, I present a new approach for *ab initio* calculations of nuclei that combines these precise and systematic forces with Monte Carlo simulation techniques that allow for exact solutions of the nuclear $A$-body problem. A short introduction of this method is given and a few assorted results concerning the spectrum and structure of $^{12}$C and $^{16}$O are presented. The framework further allows one to study the properties of nuclei in worlds that have fundamental parameters different from the ones in Nature. This allows for a physics test of the anthropic principle by addressing the question how strongly the generation of the life-relevant elements depends on the light quark masses and the electromagnetic fine structure constant.

## 1. INTRODUCTION

The nuclear many-body problem continues to be one of the most interesting and demanding challenges in contemporary physics. With the advent of FAIR, FRIB and the many existing radioactive ion beam facilities, a detailed and accurate theoretical understanding of nuclear structure and reactions is mandatory. A major breakthrough in nuclear theory was initiated through the work of Steven Weinberg [1], who made the first steps for an effective field theory (EFT) solution of the nuclear force problem. In this approach, two- and multi-nucleon forces as well as the response to external electromagnetic and weak probes can be calculated systematically, precisely and consistently. In addition, the so important issue of assigning theoretical errors can be dealt with naturally. The very hot topic of the quest for three-nucleon forces was already addressed in this journal [2] which also contained a short introduction into the framework of chiral nuclear EFT. In this scheme, the nuclear forces are given in terms of one-, two-, … pion exchanges and smeared local multi-nucleon operators, that parameterize the short-distance behaviour of the nuclear forces. These operators come with unknown coupling constants, the so-called low-energy constants (LECs) that must be determined from a fit to nucleon-nucleon scattering and a few three-body data. For systems up to four nucleons, these forces have been tested in extensive detail and scrutinized. One of the present research foci is the calculation and investigation of higher order corrections to the three-nucleon forces as well as the constuction of electroweak current operators. For reviews on the method and many results, see e.g. Refs. [3,4].

But what about larger nuclei? There are two different venues. The first one is to combine these chiral forces, eventually softened using renormalization group methods, with well developed many-body approaches like the no-core-shell model, the coupled cluster approach and so on. There have been quite a few activities in such type of approaches, see e.g. Refs. [5-10] for some recent works. Another approach, and this is the one I will discuss in what follows, is to combine the chiral forces with Monte Carlo simulation techniques, that have been successfully applied in gauge field theory (lattice QCD), condensed matter systems and other areas of physics. This novel method will be called *nuclear lattice simulations* (or nuclear lattice EFT) in the following.

In this short review, I first introduce the formalism in a very simplified manner and discuss the scope of the method, then show a few assorted physics results and finally address the question about the viability of carbon-oxygen based life on Earth when one changes certain fundamental parameters of the Standard Model that control nuclear physics.

## II. FORMALISM AND SCOPE

The basic ingredient in this framework is the discretization of space-time, see Ref. [11] for details. Space-time is represented by a grid. This lattice serves as a tool to compute the nuclear matrix elements under consideration. More precisely, one first performs a Wick rotation in time so that the time evolution operator behaves as exp$(-Ht)$, with $H$ the nuclear Hamiltonian. Space-time is then coarse-grained as shown in Fig. 1. In the three spatial directions, the smallest distance on the lattice is given by the lattice spacing $a$, so that the volume is $L \times L \times L$, with $L = N a$ and $N$ an integer, whereas in the time direction one often uses a different spacing $a_t$, and $L_t = N_t a_t$ is chosen as large as possible. Typical values are $N = 6$ and $N_t = 10...15$. As the Euclidean time becomes very large, one filters out the ground state as it has the slowest fall-off $\sim$exp$(-E_0 t)$, with $E_0$ the ground state energy. Excited states can also be investigated. This, however, requires some more effort. The nucleons are placed on the lattice sites as depicted in Fig. 1. Their interactions are given by pion exchanges and multi-nucleon operators, properly represented in lattice variables. So far, nuclear lattice simulations have been carried out using two- and three-nucleon forces at next-to-next-to-leading order (NNLO) in the chiral expansion. The Coulomb force and isospin-breaking strong interaction effects are also included, thus one has all required ingredients to describe the structure of nuclei. The lattice is used to perform a numerically exact solution of the $A$-nucleon system, where the atomic number $A = N + Z$ counts the neutrons and protons in the nucleus under investigation.

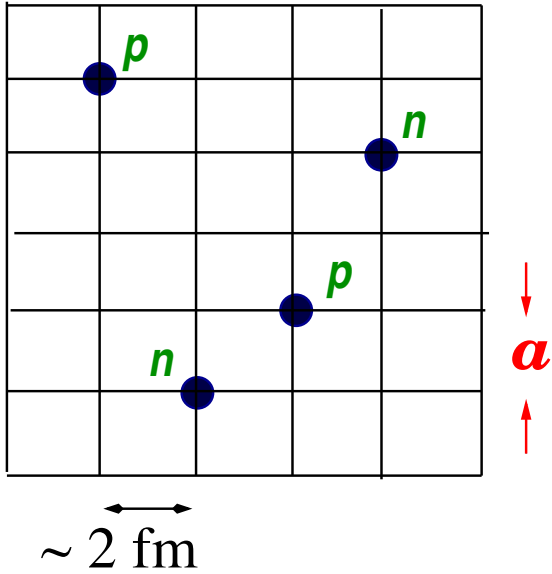

FIG. 1: *Two-dimensional representation of the space-time lattice. The smallest length on the lattice is the lattice spacing $a$, and the protons (p) and neutrons (n) are placed on the lattice sites.*

It is important to realize that the finite lattice spacing entails a maximum momentum $p_{max} = \pi/a$, so that for a typical lattice spacing of $a = 2$ fm, one has a maximum momentum of about 300 MeV, that is one deals with a very soft interaction. The main advantage of this scheme is, however, the approximate spin-isospin SU(4) symmetry of the nuclear interactions already realized by Wigner in 1937 [12]. Because of this approximate symmetry, the malicious sign oscillations that plague any fermion simulation at finite density, are very much suppressed, quite in contrast to the situation in lattice QCD. A lucid discussion of this issue has been given by Chen, Schäfer and Lee [13]. Consequently, alpha-type nuclei with $N = Z$, and spin and isospin zero can be simulated most easily. However, in the mean time our collaboration has developed a method that allows for a remarkable suppression of the remaining sign oscillations in nuclei with $N \neq Z$. One more remark on the formalism is in order. The simulation algorithms sample *all possible* configurations of nucleons, in particular one can have up to four nucleons on one lattice site. Thus, the so important phenomenon of clustering in nuclei arises quite naturally in this novel many-body approach.

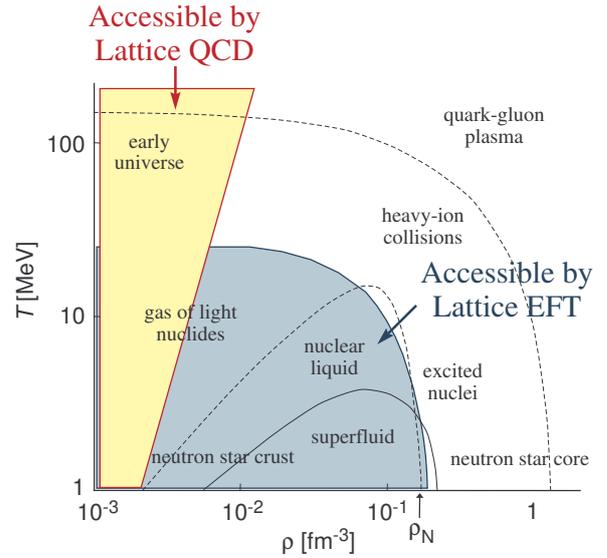

FIG. 2: *Phase diagram of strongly interacting matter. Here, $\rho$ denotes the density, with $\rho_N$ the density of nuclear matter, and $T$ is the temperature. For further details, see the text. Figure courtesy of Dean Lee.*

In Fig. 2, the phase diagram of strongly interacting matter is shown in the standard temperature versus density plot. Apart from nuclear structure research, nuclear lattice simulations can also be used to explore nuclear matter, neutron matter or other more exotic phases as indicated by the dark (blue) area. For comparison, the part of the phase diagram accessible to lattice QCD is depicted by the light (yellow) area, which is much narrower because of the sign oscillations discussed earlier. Therefore, we can systematically explore many fascinating aspects of strongly interacting matter, but for this short review I concentrate on some recent results pertinent to the structure of nuclei.

## III. ASSORTED RESULTS

Before presenting results, we must fix parameters. We have 9 LECs related to the two-nucleon force that can be fixed from the low partial waves in neutron-proton scattering. Two further LECs from isospin-breaking four-nucleon operators are fixed from the *nn* and the *pp* scattering lengths, respectively. In addition, one has two LECs appearing in the three-nucleon force, that can e.g. be fixed from the triton binding energy and the doublet neutron-deuteron scattering length. The first non-trivial prediction is then the binding energy difference of the triton and $^3$He, we find $E(^3H) - E(^3He) = 0.78(5)$ keV, in good agreement with the empirical value of 0.76 keV [14]. That we can reproduce this small effect with good accuracy gives us confidence that we have all aspects of the strong and electromagnetic forces relevant to nuclear physics under control.

If one invents a new theoretical scheme, it is absolutely necessary to solve a problem which other methods could not deal with, otherwise this new approach is not accepted easily in the community. Therefore, the first nucleus we investigated was $^{12}$C, more precisely the ground state and its low-lying even-parity excitations. The most interesting excited state in this nucleus is the so-called Hoyle state that plays a crucial role in the hydrogen burning of stars heavier than our sun and in the production of carbon and other elements necessary for life. This excited state of the carbon-12 nucleus was postulated by Hoyle [15] as a necessary ingredient for the fusion of three alpha particles to produce a sufficient amount of carbon at stellar temperatures. Without this excited state that is located very close to the $^4$He+$^8$Be threshold (thus leading to a resonant enhancement of the production rate), much too little carbon would be generated and consequently also much too little oxygen, thus making life on Earth impossible. Although the Hoyle state was seen experimentally more than a half century ago [16], nuclear theorists have tried unsuccessfully to uncover the nature of this state from first principles. Using nuclear lattice simulations, we could perform an *ab initio* calculation of this elusive state [17]. Here, by *ab initio* we mean that all parameters appearing in the nuclear forces have been determined in the two- and three-nucleon systems and that the 12 particle problem has been solved numerically exactly using Monte Carlo techniques. The resulting spectrum of the lowest states with even parity is shown in Fig. 3. One observes a nice agreement between theory and experiment. Not only does one get the Hoyle state at its correct position but also the much investigated $2^+$ excitation a few MeV above it. Further insight into the structure of the Hoyle state was obtained in Ref. [18], where the structure of these states was investigated. In all these states, alpha clustering plays a prominent role. For the ground state and the first excited $2^+$ state, one finds a compact triangular configuration of alpha clusters. For the Hoyle state and its $2^+$ excitation, however, the dominant contribution is a „bent arm" or obtuse triangular alpha cluster configuration. A remaining challenge is the calculation of radii and transition moments beyond leading order so as to make contact to the precise measurements of electromagnetic observables performed e.g. at the S-DALINAC in Darmstadt [19].

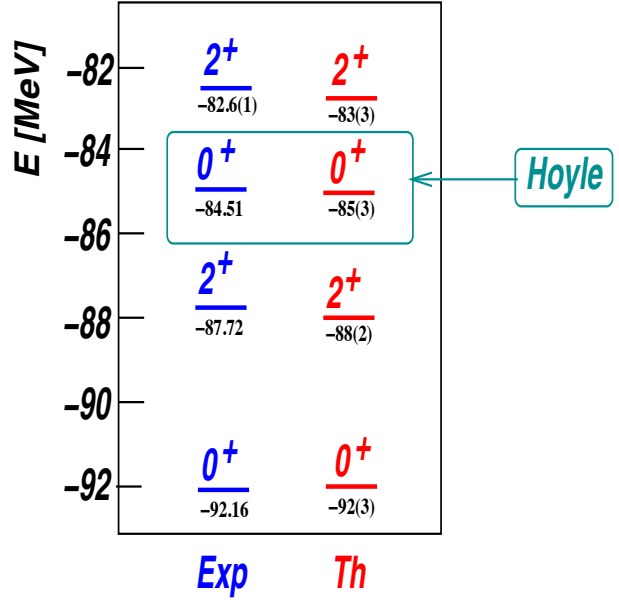

FIG. 3: *Even-parity spectrum of $^{12}$C. On the left, the empirical values are shown, whereas the right column displays the nuclear lattice simulation results from Ref. [16].*

Another alpha cluster-type nucleus is $^{16}$O, that also plays an important role in the formation of life on Earth. Since the early work of Wheeler [20], there have been several theoretical and experimental works that lend further credit to the cluster structure of $^{16}$O. However, no *ab initio* calculations existed that gave further support to these ideas. This gap was filled in Ref. [21] where nuclear lattice simulations have been used to investigate the low-lying even-parity spectrum and the structure of the ground and first few excited states. It is found that in the spin-0 ground state the nucleons are arranged in a tetrahedral configuration of alpha clusters, cf. Fig. 4. For the first $0^+$ excited state, the predominant structure is given by a square arrangement of alpha clusters, as also shown in Fig. 4. There are also rotational excitations of this square configurations that include the lowest $2^+$ excited state. These cluster configurations can be obtained in two ways. First, one can investigate the time evolution of the various cluster configurations shown in Fig. 4 and extract e.g. the corresponding energies as the Euclidean time goes to infinity. Second, one can also start with initial states that have no clustering at all. One can then measure the four-nucleon correlations. For such initial states, this density grows with time and stays on a high level. For the cluster initial states, these correlations start out at a high level and stay large as a function of Euclidean time. This is a clear indication that the observed clustering is not build in by hand but rather follows from the strong four-nucleon correlations in the $^{16}$O nucleus.

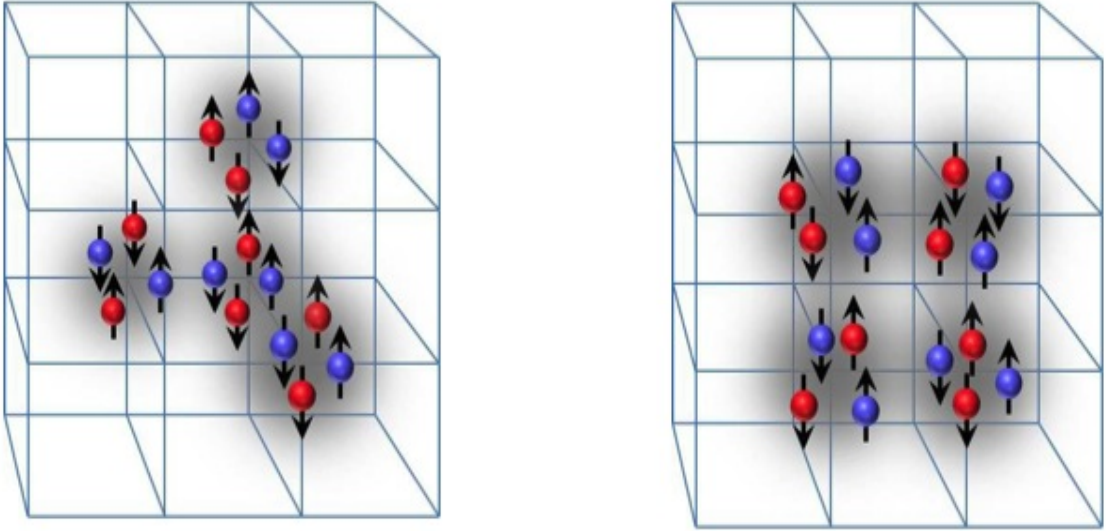

FIG. 4: *Schematic illustration of the alpha cluster states in the tetrahedral (left panel) and the square (right panel) configuration.*

## IV. ANTHROPIC CONSIDERATIONS

Another fascinating aspect of this method is that it allows to test the changes of the generation of the life-relevant elements under variations of the fundamental constants of the Standard Model. For the case of nuclear physics, these are the light quark masses and the electro-magnetic fine structure constant $\alpha_{EM}$. While the light quarks generate only a small contribution to the mass of the nucleon - showing that mass generation is not entirely given by the Higgs boson - their masses are comparable to the typical binding energy per nucleon. Therefore, variations in the quark masses will lead to changes in the nuclear binding. In fact, nuclear binding appears to be fine-tuned by itself. A deeper understanding of this particular fine-tuning in Nature has so far been elusive.

The already discussed Hoyle state has often been heralded as a prime example of the *anthropic principle*, that states that the possible values of the fundamental parameters are not equally probable but take on values that are constrained by the requirement that life on Earth exists. A crucial parameter in the triple-alpha reaction rate is the difference between the Hoyle state energy and three times the alpha nucleus mass, $\Delta E = 380$ keV. Already in 2004, Schlattl et al. [22] showed that $\Delta E$ could be changed by about ±100 keV so that one still produces a sufficient amount of carbon and oxygen in stars. This is a 25% modification that does not appear very fine-tuned. But how does this translate into the fundamental parameters of the Standard Model, i.e. into changes of the light quark mass and the fine structure constant? This can be answered by employing the chiral EFT approach to the nuclear forces. In fact, as in QCD the pion mass squared is proportional to the sum of the light up and down quark masses, we simply need to study the variation of the forces under changes of the pion mass. This has been done to NNLO in the chiral expansion in Ref. [23], where also constraints on quark mass variations from the abundances of the light elements from the Big Bang were derived. Armed with that, in Refs. [24] a detailed study of the resonance condition in the triple alpha process was performed, leading to the conclusion that quark mass variations of 2...3% are not detrimental to the formation of life on Earth. However, this number is afflicted with a sizeable uncertainty that can eventually be overcome from lattice QCD studies of the nucleon-nucleon scattering lengths. It could also been shown that the various fine-tunings in the triple alpha process are correlated, as had been speculated before [25]. Further, the possible variation of the fine structure constant can be also derived, changes in $\alpha_{EM}$ of ±2.5% are consistent with the requirement that $\Delta E$ changes by at most 100 keV in magnitude.

Consequently, the light quark masses and the fine structure constant are fine tuned. Beyond these rather small changes in the fundamental parameters, the anthropic principle appears necessary to explain the observed abundances of $^{12}$C and $^{16}$O. For a recent review on the applications of the anthropic principle, the reader is referred to Ref. [26].

## V. SUMMARY AND CONCLUSIONS

In this article, I have given a short review about the novel method of nuclear lattice simulations and showed some first promising results for nuclei. In addition, an application to testing the anthropic principle, which has consequences much beyond nuclear physics, was discussed. Of course, there remains much work to be done. In particular, the removal of lattice artefacts through the finite lattice spacing and the finite volume has to be further improved, see e.g. the recent work on the restoration of rotational symmetry for cluster states [27]. In addition, the methods to reduce the errors in the extraction of the signals from the Euclidean time interpolation have to be sharpened, at present ground state energies of nuclei up to $A = 28$ can be extracted with an accuracy of 1% or better [28]. The computing time scales approximately as $A^2$, so that heavier nuclei can also be investigated in the future. Further, the underlying forces have to be worked out and implemented to NNNLO in the chiral expansion. Another line of research is to allow for the continuum limit $a \to 0$ by formulating the EFT with a cut-off on the lattice. A first step was done in Ref. [29]. Another area of research concerns the equation of state of neutron and nuclear matter and the possibility of various pairing phenomena in such dense systems. Finally, preliminary investigations towards the inclusion of hyperons are also done to tackle problems in hypernuclear physics. There is a bright future for applying and improving nuclear lattice simulations and I would like to see more groups employing this powerful tool.

## Acknowledgements

I would like to thank all members of the NLEFT collaboration for sharing their insight into the topics discussed here, especially Dean Lee, Timo Lähde and Evgeny Epelbaum. I would also like to thank Evgeny Epelbaum and Dean Lee for comments on the manuscript and for help with this strange type-setting system. Computational resources were provided by the Jülich Supercomputer Center and the RWTH Aachen. Work supported in part by the BMBF, the DFG, the HGF, the NSFC and the EU.